\documentclass[reprint,aps]{revtex4-1}
\usepackage{graphicx}
\usepackage{subfigure}
\usepackage{float}
\usepackage{color}
\usepackage{hyperref}
\begin{document}

\title{Stability aspects of relativistic thin magnetized disks}

\author{Vanessa P. de Freitas}\email{vpachecof@gmail.com}
\affiliation{Centro Brasileiro de Pesquisas F\'\i sicas, 22290-180 Rio de Janeiro, RJ, Brazil.}
\author{Alberto Saa}\email{asaa@ime.unicamp.br}
\affiliation{Departmento de Matem\'atica Aplicada, 
 IMECC--UNICAMP,  13083-859 Campinas, SP, Brazil.}

\begin{abstract}
We adapt the well known ``displace, cut and reflect'' method to construct exact solutions of the Einstein-Maxwell equations corresponding   to  infinitesimally  thin disks of matter endowed with   dipole magnetic fields, which are entirely  supported by surface polar currents on the disk. Our starting point is the Gutsunaev-Manko   axisymmetric solution describing  massive    magnetic dipoles in General Relativity, from which we obtain a continuous three-parameter family of asymptotically flat static magnetized disks with finite mass and energy. For strong magnetic fields, the disk surface density profile   resembles  some well known self-gravitating ring-like structures. 
 We show that many of these solutions  can be indeed stable and, hence, they  could be   in principle useful for the study of the abundant astrophysical situations involving  disks of matter  and magnetic fields. 
\end{abstract}

\maketitle

\section{Introduction}
Axially symmetric solutions of the Einstein equations corresponding to thin disks of matter   are rather common in the physical literature. They can be static or stationary, with or without radial pressure, accommodate  heat flow, electric charge, and halos, 
among many other possibilities, 
 see, for instance 
\cite{BonnorSackfield,MorganMorgan1,MorganMorgan2,LyndenBellPineault,
LetelierOliveira,Lemos,BLBK,LemosLetelier1,LemosLetelier2,
GonzalezEspitia,GarciaReyesGonzalez,BicakLedvinka,GonzalezLetelier1,
GonzalezLetelier2,VogtLetelier1,VogtLetelier2}.
 Many of the known thin disk solutions can be obtained by using the so-called ``displace, cut and reflect'' (DCR) method, initially  due to Kuzmin \cite{DCR}, which, in fact, has   Newtonian origin and will be briefly presented in the next section. Matter disks and other
 self-gravitating structures in the presence of magnetic fields are particularly relevant due to the abundance of possible astrophysical applications, and there are indeed some previous examples of these solutions in the literature 
 \cite{Letelier,GutierrezGonzalez,GutierrezGarciaGonzaelez,Lee}.

In the present work, we will show how to adapt the standard  DCR method   to generate consistent solutions of the Einstein-Maxwell equations corresponding to thin disks of matter with magnetic fields entirely supported by polar surface currents on the disk, a situation clearly mimicking  realistic astrophysical thin plasma disks \cite{Plasma}. 
Our starting point will be the 
 Gutsunaev-Manko two-parameter family of  solutions  describing massive axisymetric objects endowed with a magnetic dipole moment in General Relativity \cite{GutsunaevManko1, GutsunaevManko2, GutsunaevManko3}. In contrast with other solutions with magnetic fields, as for instance the long standing and well known one  due to Bonnor \cite{Bonnor}, for the  Gutsunaev-Manko family, the object mass and dipole magnetic moment are really independent quantities and, hence, we are able to generate   a continuous three-parameter family of magnetized disks with finite total mass and energy.
 The three parameters can  determine    univocally, for instance, the disk 
 total mass,  dipole magnetic moment,  and central superficial density. 
Such parameters  
 can be chosen to achieve some desired physical properties, with special emphasis, of course, to the stability of the solution. 
For the  stability analysis of the disk, we consider, besides the 
  usual  generalized radial Rayleigh criteria\cite{Rayleigh, Letelier2},  also the vertical stability (oblique orbits)\cite{Ronaldo}.  
  We show that one indeed has a large continuous family of magnetized disks satisfying these stability criteria, which  could be useful, in principle, for the study of the abundant astrophysical situations involving  disks of matter  and magnetic fields. Our magnetized disks have only azimuthal pressure, and hence they can be interpreted physically as composed by counter-rotating particles, see, for instance, \cite{GonzalezEspitia} for further references on this rather common hypothesis for static disks configurations in General Relativity. 
Incidentally,   we notice   that there are indeed some recent observational evidences for counter-rotating stellar disk, see \cite{CR,CR1} and references therein.

This article is organized as follows. Section \ref{Disks} provides an overview of the DCR method and its necessary adaptation to generate viable magnetized disks. 
For sake of completeness, we also review briefly the main pertinent results about the stability of thin disks. Section III is devoted to construct and discuss the stability of the new solutions, and the last section is left for some concluding remarks about the stability of our disks.

\section{Relativistic thin Disks}
\label{Disks}

\begin{figure}[h]
\center
\includegraphics[scale=0.45]{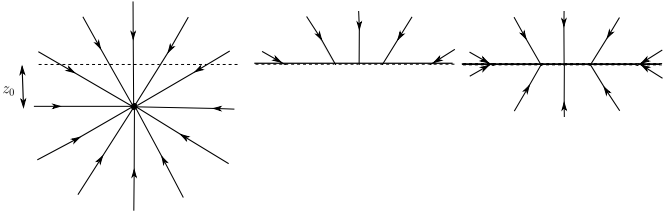}
\caption{Illustration of the   ``displace, cut and reflect'' (DCR) Kuzmin method starting from a central field in cylindrical coordinates. First, one choses a hypersurface (dashed line) separating the spacetime in two regions (left), then we discard the part containing the source singularity (middle), and finally reflect the field on the hypersurface (right).}
\label{cdr}
\end{figure}

The Kuzmin ``displace, cut and reflect'' (DCR) method \cite{DCR} can be used to generate  rather generic thin disk configurations starting from a given solution of Einstein (or even Laplace) equations. Considering, for instance, a point-source solution,  we can divide this method accordingly to the following steps: first we choose a hypersurface ($z=z_0$ in cylindrical coordinates, in our case, see Fig. \ref{cdr})
 separating the spacetime in two regions,   one of them containing  the source of the gravitational field. Second, we discard the part that contains the source and, third, use the hypersurface $z=z_0$ to reflect the remaining  part. 
The result of the application of this procedure will be generically a regular spacetime with a surface singularity or, in other words,
a solution of Einstein equations with an infinite thin disk of matter located at $z=z_0$. Despite of being infinite, in many cases, as in the present one, these disks have finite
total mass and energy and, hence, they might be considered as approximations to some finite size disks.  
 The DCR procedure is fully equivalent to make the following mathematical transformation  \cite{VogtLetelier1} 
\begin{equation}
\label{dcr}
z\rightarrow |z| + z_0
\end{equation} 
for all the  pertinent quantities, 
where the disk plane now will be given   by $z=0$ and $z_0>0$ is a free parameter. Such a method can be applied virtually to any gravitational  solution, relativistic or Newtonian, resulting generically in gravitational fields supported by surface distributions of matter, see \cite{DCR1}  for further references.

We are mainly interested here in axisymmetric static solutions of the Einstein equations, and these solutions can be conveniently 
described in   cylindrical coordinates $(t,r,z,\varphi)$ as
\begin{eqnarray}
\label{geral}
ds^2 &=& g_{\mu\nu}dx^\mu dx^\nu \\
&=&e^{\psi(r,z)}dt^2- e^{\eta(r,z)} \left( dr^2+dz^2\right)- r^2e^{\gamma(r,z)} d\varphi^2.\nonumber
\end{eqnarray}
The application of (\ref{dcr}) gives origin generically   to functions of class $C^0$ across the disk plane. Using the standard notation $[f] = \left.f\right|_{z=0^+} - \left.f\right|_{z=0^-}$ for the discontinuities across  the hypersurface $z=0$, one has $
\left[g_{\mu \nu}\right] = 0
$. On the other hand, the $z$ derivative of the metric tensor is typically discontinuous at
$z=0$, and the quantities 
\begin{equation}
\label{derivada_descontinuidade}
 \left[\partial_z g_{\mu \nu }\right] = b_{\mu \nu} 
\end{equation}
will determine all the physical and geometrical properties of the disk. In particular, 
the Christofel symbols and the Riemann curvature tensor $R^{\sigma}_{\ \alpha \gamma\beta}$ can be define by means
of distributions involving (\ref{derivada_descontinuidade}), leading to \cite{Taub}
\begin{equation}
R^{\sigma}_{\ \alpha \gamma\beta}= {\cal R}^{\sigma}_{\ \alpha \gamma\beta} +H^{\sigma}_{\ \alpha \gamma\beta}\hat\delta(z),
 \end{equation}
 where $\hat\delta(z)$ stands for the covariant Dirac $\delta$-function \cite{Ronaldo}, ${\cal R}^{\sigma}_{\ \alpha \gamma\beta}$ is the (smooth) Riemann curvature tensor for $z\ne 0$, and
  \begin{equation}
 \label{H}
 H^{\sigma}_{\ \alpha \gamma\beta}= \frac{e^{ \eta/2}}{2}\left( \delta^z_{\alpha}\delta^z_{\gamma}b^{\sigma}_{\beta}-g^{z\sigma}\delta^ z_{\gamma}b_{\alpha\beta}-\delta^z_{\alpha}\delta^z_{\beta}b^{\sigma}_{\gamma} +g^{z\sigma}\delta^z_{\beta}b_{\alpha\gamma}  \right).
\end{equation}
From the contractions of (\ref{H}), one can calculate directly the disk energy momentum tensor $Q_{\alpha\beta}$, which
will be given by 
\begin{equation} 
\label{H_q}
H_{\alpha\beta}-\frac{1}{2}g_{\alpha\beta}H=8\pi Q_{\alpha\beta},
\end{equation}
with $H_{\alpha\beta} = H^{\sigma}_{\ \alpha\sigma\beta}$ and $H=H^{\gamma}_{\ \gamma}$. 
One   finally has \cite{Ronaldo}
\begin{eqnarray}
\label{Qab}
Q^\alpha_\beta &=&\frac{e^{\eta/2}}{16\pi}\left[b^{z\alpha}\delta^z_\beta b^{zz}\delta^\alpha_\beta + g^{z\alpha}b^z_\beta - g^{zz}b^\alpha_\beta \right. \nonumber \\ & &+\left. b^\sigma_\sigma(g^{zz}\delta^\alpha_\beta - g^{z\alpha}\delta^z_\beta)\right],
\end{eqnarray}
which is diagonal for metrics of the type (\ref{geral}). Its components 
correspond to the surface energy density and
the pressures in the radial,  axial, and azimuthal  directions, respectively,
\begin{equation}
\label{grandezas}
Q^\alpha_\beta = {\rm diag}(
\sigma ,
- P_r, - P_z, - P_{\varphi}).
\end{equation}
Moreover, is clear from (\ref{Qab}) that $P_z=0$ for our case, as one would indeed naturally expect for infinitesimally thin disks.  Assuming  the system to be symmetric under reflections $z\to-z$, one can calculate
$\sigma$, $P_r$, and $P_\varphi$ for static axisymmetric spacetimes with metric (\ref{geral}),
leading to \cite{Ronaldo}
\begin{eqnarray}
\label{sigma}
\sigma &=& -\frac{e^{-\eta/2}}{8\pi} \left(\frac{\partial \gamma}{\partial |z|} + \frac{\partial \eta}{\partial |z|}  \right)_{z=0}, \\
\label{Pr}
P_r &=& \frac{e^{-\eta/2}}{8\pi} \left(\frac{\partial \psi}{\partial |z|} + \frac{\partial \gamma}{\partial |z|}  \right)_{z=0} ,\\
\label{Pvarphi}
P_\varphi &=& \frac{e^{-\eta/2}}{8\pi} \left(\frac{\partial \psi}{\partial |z|} + \frac{\partial \eta}{\partial |z|}  \right)_{z=0} ,
\end{eqnarray}
where the notation $\frac{\partial} {\partial |z|}$ means that the derivative is calculated after the substitution (\ref{cdr}).
We are now in conditions to formulate the stability criteria we will use for the disk. 

However, before starting the stability discussion, it is important to stress that the use of (\ref{cdr}) alone is not enough to generate viable magnetized disks. The problem is that in the same way the DCR method induces a superficial density of matter at $z=0$, it will generically do with a superficial density of magnetic monopoles on the disk  \cite{monopolos}, 
jeopardizing any possible realistic physical application for these solutions. Fortunately, this can be easily fixed. Suppose we have a solution of the Einstein-Maxwell equations of the form (\ref{geral}) with a magnetic   field given by the 
electromagnetic quadri-potential 
\begin{equation}
\label{quadri}
A_\mu = \left[0, 0 , 0, A_\varphi(r, z)\right].
\end{equation}
The linearity of the Maxwell equations and the quadratic structure of the Maxwell energy-momentum tensor imply that both $A_\mu$ and $-A_\mu$ are solutions of the Einstein-Maxwell equations for the same metric tensor (\ref{geral}). One can avoid the appearance of magnetic monopoles on the disk if we combine  (\ref{cdr})  with the transformation
\begin{equation}
\label{cdr1}
A_\mu \to ({\rm sgn}\, z) A_\mu .
\end{equation} 
By using (\ref{cdr}) and (\ref{cdr1}), we will be able to generate  disks of matter with dipole magnetic   fields entirely supported by surface polar currents on the disk. Since the metric tensor is invariant under (\ref{cdr1}), the stability analysis, based solely in the energy-momentum tensor of the disk, is not affected by this transformation. On the other hand, (\ref{cdr1}) is crucial  to obtain the polar currents which indeed generate the disk magnetization.

\subsection{Disk stability conditions}

As we will see, the matter content of our disks has no radial pressure, and in this case the
disk   is usually viewed as being composed  of counter-rotating identical particles, which guarantees, besides the vanishing of the radial pressure, a vanishing total angular momentum for the disk, 
despite the rotation of its matter content, see \cite{GonzalezEspitia} for further details
and references. 
We will perform two stability tests for  our solutions: radial  and vertical perturbations.  
In fact, they consist   in the (linear) stability analysis of the circular orbits at $z=0$ against small perturbations, and both tests can be derived from the geodesic motion on the disk plane and around its vicinity. 
 Since (\ref{geral}) is static and axisymmetric, its geodesic equations will have two independent constants of motion: the total energy $H$ and the angular momentum around the $z$ direction $L_z$. The (reduced) Hamiltonian for the geodesics of the metric (\ref{geral}) is given by \cite{Ronaldo}
\begin{equation}
\label{Ham}
H^2(r,z,p_r,p_z) = e^{\psi-\eta}\left(p_r^2 + p_z^2 \right) + V_{\rm eff}(r,z),
\end{equation}
where the effective potential 
\begin{equation}
\label{Veff}
V_{\rm eff}(r,z) = e^\psi\left(1+ e^{-\gamma}\frac{L_z^2}{r^2} \right) 
\end{equation}
will be crucial for both stability tests. 

A pertinent and rather deep question is if this kind of stability analysis of the circular geodesics in the disk plane would be enough to guarantee the disk stability beyond the counter-rotating hypothesis. A full answer to this question is obviously  out of scope of the present paper, but we will return to this point in the last section.

\subsubsection{The Rayleigh criterion}

The Rayleigh criterion establishes the requirements for the circular orbits on the disk    be stable against radial perturbations. Its relativistic version has been intensively studied in recent years, see \cite{Ronaldo}. 
Essentially, it can be deduced from the behavior of circular planar solutions of (\ref{Ham}), {\em i.e.},  the solutions with $z=p_z=p_r=0$ and, consequently, $r=\bar r$ constant, which,
of course, occurs for  $\partial_r V_{\rm eff}(\bar r,0) =0 $. The stability of such circular orbits requires
\begin{equation}
\label{Ray}
\frac{\partial^2 V_{\rm eff}}{\partial r^2} = \frac{\partial_r e^\psi}{L_z^2}
\partial_r L^2_z > 0
\end{equation}
for all $r$, where
\begin{equation}
L_z^2  = \frac{r^4e^{2\gamma} \partial_r e^\psi}{e^\psi \partial_r (r^2 e^\gamma) - r^2 e^\gamma \partial_r e^\psi}  
\end{equation}
 is evaluated at $z=0$. As one can see, the disk radial stability   requires that both $e^\psi$ and $L_z^2$ be monotonically increasing functions of $r$ on the disk plane $z=0$. The situation here is clearly analogous to the Newtonian case \cite{DCR,Vieira2014}. 
 
 \subsubsection{Vertical stability}

Since the dynamics of the geodesics are in fact singular at $z=0$, the stability of the oblique
orbits will be determined solely by the ``restoring'' vertical force $\partial_{|z|} V_{\rm eff}$ \cite{Ronaldo}.
In fact, for the metric (\ref{geral}), the condition 
\begin{equation}
\label{vertical}
\frac{\partial V_{\rm eff}}{\partial |z|} = 
4\pi e^{\psi+\eta/2}\left[ P_r + \left( 1+ \frac{2L^2_z}{r^2e^\gamma}\right) \left( \sigma+P_\varphi\right)\right]
 > 0,
\end{equation}
evaluated on the plane $z=0$, 
 is enough to guarantee the vertical stability of circular orbits in our disks.

\section{Magnetized  disks}

The Gutsunaev-Manko spacetimes \cite{GutsunaevManko1, GutsunaevManko2, GutsunaevManko3} form a
large family of asymptotically flat,  axisymmetric, and stationary
 solutions of the Einstein-Maxwell equations. We are mainly interested here in one of their simplest
 sub-cases, the  continuous two-parameter family corresponding to the gravitational field of massive
magnetic dipoles \cite{GutsunaevManko2}. It can be conveniently written,  using the cylindrical coordinates (\ref{geral}), in the   Weyl line element form, for which 
\begin{equation}
\label{gs1}
e^\psi = e^{-\gamma} =
  \frac{x-1}{x+1} f^2,
\end{equation}
and
\begin{equation}
e^\eta =  \frac{(x+1)^2}{x^2-y^2}  \left(\frac{g}{f}\right)^2,
\end{equation}
with
\begin{equation}
\label{f}
f =  \frac{(x^2-y^2+\alpha^2(x^2-1))^2+4\alpha^2x^2(1-y^2)}{(x^2-y^2+\alpha^2(x-1)^2)^2-4\alpha^2y^2(x^2-1)} 
\end{equation}
and
\begin{equation}
\label{g}
g = \left(\frac{(x^2-y^2+\alpha^2(x^2-1))^2+4\alpha^2x^2(1-y^2)}{(1+\alpha^2)^2(x^2-y^2)^2}\right)^2,
\end{equation}
where $\alpha$ is a dimensionless constant, and $x$ and $y$ are the usual prolate coordinates, related to the cylindrical coordinates $r$ and $z$ by
\begin{eqnarray}
\label{t1}
x &=& \frac{1}{2k}\left(\sqrt{r^2+(z+k)^2} + \sqrt{r^2+(z-k)^2}  \right), \\
\label{t2}
y &=& \frac{1}{2k}\left(\sqrt{r^2+(z+k)^2} - \sqrt{r^2+(z-k)^2}  \right),
\end{eqnarray}
where $k$ is a dimensional constant. 
The dipole magnetic field, which has the form (\ref{quadri}), is given by
\begin{eqnarray}
\label{quadri2}
 A_\varphi &=& \frac{4k\alpha^3}{1+\alpha^2}(1-y^2)\times \\ 
&& \frac{2(1+\alpha^2)x^3+(1-3\alpha^2)x^2+y^2+\alpha^2}{(x^2-y^2+\alpha^2(x^2-1))^2+4\alpha^2x^2(1-y^2)}.\nonumber
\end{eqnarray}
In order to interpret the physical role of the two parameters $k$ and $\alpha$, one can cast the metric in spherical coordinates and analyze the asymptotically flat limit \cite{GutsunaevManko1, GutsunaevManko2, GutsunaevManko3}, leading to the 
conclusion that these parameters  are related to the mass $m$ and magnetic dipole $\mu$ of the central object by the expressions
\begin{eqnarray}
\label{mass}
m &=& \frac{1-3\alpha^2}{1+\alpha^2}k, \\
\label{mu}
\mu &=& \frac{8\alpha^3}{(1+\alpha^2)^2}k^2.
\end{eqnarray} 
The only restrictions on these parameters in order to have physically meaningful solutions, for
our purposes,  are $k > 0$ and  $ 3\alpha^2 < 1$.   The Schwarzschild case is recovered for $\alpha = 0$, and the solution can accommodate any values for the mass $m$ and the dipole magnetic moment $\mu$. In particular, $\alpha$ is given by
\begin{equation}
\frac{8\alpha^3}{(1-3\alpha^2)^2} = \frac{\mu}{m^2},
\end{equation}
from where it is clear that with $3\alpha^2<1$ one can effectively cover all possibilities for
$\mu$ and $m$.
The parameter $\alpha$ is clearly dimensionless,  $k$ is the only massive parameter for these solutions. All the dimensional quantities in our analysis will be always expressed in units of $k$. The Gutsunaev-Manko solutions are asymptotically flat, and such a crucial property will be inherited by our disks. 

The magnetized disks are generated by doing  the transformation (\ref{cdr}) and 
simultaneously $\alpha\to-\alpha$, which implements (\ref{cdr1}), in all  pertinent expressions, giving origin to a three-parameter $(k,\alpha,z_0)$ family of disk solutions for
the Einstein-Maxwell equations. We will restrict ourselves to configurations
such that $z_0>k$ in order to avoid the rather intricate causal structure of the Gutsunaev-Manko solution near its center, otherwise
it would be impossible to have stable disk configurations. We will return to this point in the last section.  Despite all pertinent expressions here involve essentially only rational  functions and, hence, all the calculations can be straightforwardly carried out, the resulting expressions are quite cumbersome, and so we opt to present our results graphically. Nevertheless, we present in the Appendix all the necessary details to evaluate the algebraic expressions.
The explicit expressions for the Gutsunaev-Manko solutions allow some useful simplifications. In particular, since $\psi = -\gamma$, see (\ref{gs1}), one gets from (\ref{Pr}) that, as we have already
advanced,
$P_r=0$, reducing the vertical stability criterion to
\begin{equation}
\sigma + P_\varphi = \frac{e^{-\psi-\eta/2}}{4\pi}  \frac{\partial e^\psi}{\partial |z|}    > 0,
\end{equation} 
at $z=0$, or, in other words, the disk must obey the null energy condition to assure
the stability of vertical perturbations (oblique orbits). Hence, we will guarantee both the
 radial  and vertical  stabilities 
provided that,  on the disk plane $z=0$, 
 $\partial_{|z|}e^\psi$ is a positive   function, and $e^\psi$    and
\begin{eqnarray}
\label{l2}
 L^2_z = \frac{r^3\partial_re^\psi}{2 e^{2\psi} - r\partial_re^{2\psi}}
\end{eqnarray} 
are monotonically increasing in $r$,    see (\ref{Ray}).  

\begin{figure}[h]
\includegraphics[scale=0.4]{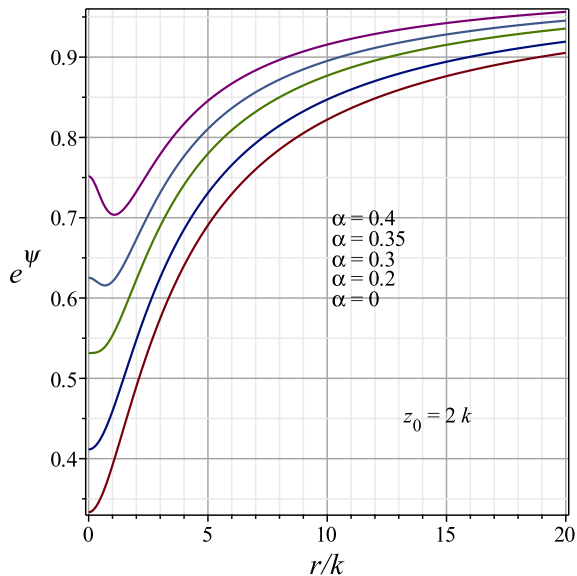}
\caption{Typical aspect of the function  $e^\psi$ given by (\ref{gs1}) on the disk  plane $z=0$, here depicted for $z_0=2k$ and several values of
$\alpha$, with the respective curves from bottom to top.  The non-monotonic behavior is associated with the concavity change of the function $e^\psi$ at the origin, see the text.}
\label{epsi}
\end{figure}

Let us start with the  function $e^\psi$ given by (\ref{gs1}).  It is indeed monotonically increasing in $r$ on the disk plane $z=0$ for large ranges of the parameters.  The asymptotic behavior of $e^\psi$, see (\ref{asy2}) in the Appendix, assures its monotonic character for large $r$, for any value of the parameters.  
Fig. \ref{epsi} depicts some typical cases around the disk central region. 
As one can see, the function fails to be monotonically increasing for some larger values of $\alpha$. For a given $z_0$, one needs to restrict $\alpha$ to a certain interval $(-\alpha_*,\alpha_*)$ in order to have a monotonic $e^\psi$ . We can check that $\partial_re^\psi=0$ at $r=z=0$ for all values of the parameters $\alpha$ and $z_0$. The non-monotonic phase is associated with the concavity change of the function $e^\psi$ at the origin. In order to determine the value of $\alpha_*$ as a function of $z_0$ associated with this concavity change, one needs to find the roots of a cumbersome higher order polynomial, so again we employ numerical and graphical analyses. (See the Appendix for the details on the algebraic expressions.)
\begin{figure}[h]\includegraphics[scale=0.4]{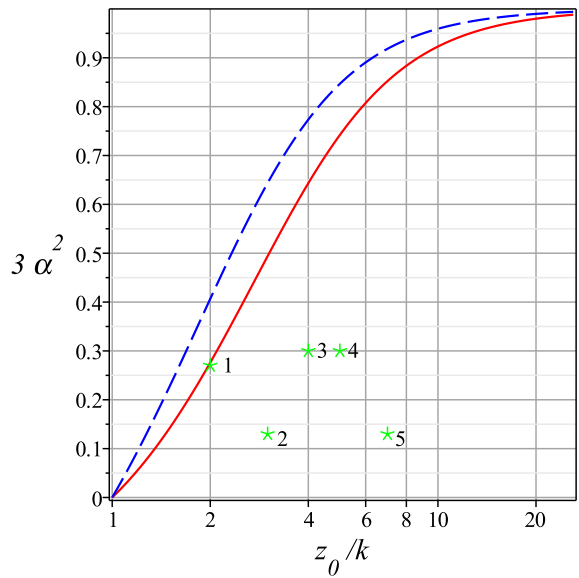}
\caption{Diagram  $3\alpha^2$ versus  $z_0/k$ (mono-log scale). The red (solid) line corresponds to
the threshold $\alpha_*$ for monotonically increasing functions $e^\psi$, see Fig. \ref{epsi}.
 Only points lying below this curve are allowed for stable circular obits on the disk. The blue (traced) line corresponds to the values of $\alpha$ and $z_0$ such that the disk central density $\sigma_0$ vanishes. In order to have a positive $\sigma_0$, the parameters $\alpha$ and $z_0$ must lay bellow this curve. As one can see, is is enough to have a  monotonically increasing  $e^\psi$ to assure a positive $\sigma_0$. The green numbered points correspond to some typical disk configurations presented in  the next figures and discussed in the text. Their  $(z_0/k, \alpha)$ values  are, respectively: $(2,0.30 )$, $(3,0.21)$, $(4, 0.32)$, $(5,0.32)$, and $(7,0.21)$. }
\label{conc}
\end{figure}
Fig. \ref{conc} shows the interval limit $\alpha_*$ as a function of $z_0$. Viable disks must have
the parameters $(z_0,\alpha)$ lying  below the depicted red (solid) curve.

The parameter $z_0$ can be related to the disk surface density, which in our case is
given by
\begin{equation}
\label{sigmad}
\sigma = \frac{e^{-3\eta/2-\psi}}{8\pi} \left( e^\eta\frac{\partial e^\psi}{\partial |z|} -e^\psi\frac{\partial e^\eta}{\partial |z|}  \right).
\end{equation}
Let us consider the central ($r=0$) disk density $\sigma_0$, which one can evaluate directly as
\begin{equation}
\label{sigma0}
\sigma_0 = \frac{1}{2\pi k}\sqrt{\frac{z_0-k}{z_0+k}}\left(
  f_x^{(0)}  + \frac{k^2f^{(0)}}{z_0^2-k^2}  
 \right) ,
\end{equation}
where the ${}^{(0)}$ superscript indicates that the corresponding quantities are calculated at the center of the disk.
 This calculation is rather tedious, but straightforward, the details are in the appendix. For fixed $\alpha$ and large $z_0$, one has
\begin{eqnarray}
\sigma_0\sim \frac{m k}{2\pi z_0^2},
\end{eqnarray}
where $m$ is the mass given by (\ref{mass}). Hence, the central surface density $\sigma_0$ decreases when $z_0$ increases  for fixed $\alpha$, {\em i.e.}, the disk becomes fainter, as one would indeed expect from the DCR method applied for an asymptotically flat spacetime. The problem, however, is that for a fixed $z_0$, there are values of $\alpha$ such that $\sigma_0<0$, challenging the idea that the disk would be formed by some kind of ordinary
counter-rotating matter. One can determine the values of $\alpha$ such that, for a given $z_0$, 
the term between parenthesis in (\ref{sigma0}) change its signal. The results are also depicted in Fig. \ref{conc}. It is important to stress that the values of $\alpha$ which leads to a negative $\sigma_0$ are always larger than those ones   assuring a monotonic function $e^\psi$, see Fig. \ref{epsi}. Hence, a monotonically increasing $e^\psi$   will also guarantee a positive central disk superficial density $\sigma_0$. Furthermore, from (\ref{sigmad}),  we  have for large $r$ 
\begin{eqnarray}
\sigma \sim \frac{mz_0}{2\pi r^3},
\end{eqnarray}
and, consequently, the total mass of the disk will be always finite. 

\begin{figure}[h]
\includegraphics[scale=0.4]{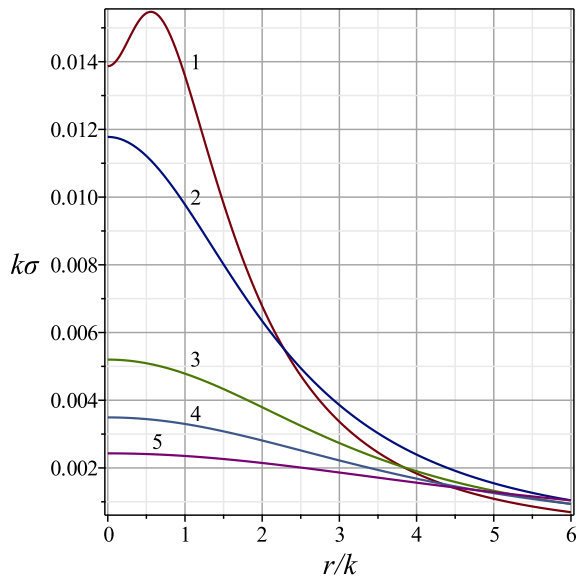}
\caption{Disk surface density $\sigma$ given by (\ref{sigmad}) as a function of $r$, for the disk configurations described in Fig. \ref{conc}. For strong magnetic fields (larger $\alpha$, as for the case of the configuration number 1), the maximum of the surface density is located at a $r>0$, suggesting a ring-like distribution.  }
\label{Fig4}
\end{figure}

It is instructive also to inspect the graphics of the disk surface energy density (\ref{sigmad}), see
Fig. \ref{Fig4}.
 For configurations near the concavity threshold (red (solid) line in Fig. \ref{conc}), as for instance the configuration number 1, the maximum density of the disk is not 
located at its center, but at a certain radius $r_*>0$, typically smaller than $z_0$. This kind of configuration resembles clearly some well known self-gravitating ring structures, see \cite{Polish} for references. In our case, such rings require strong magnetic fields to exist. For 
weak magnetic fields (small $\alpha$), the
maximum of the surface density will be always located at the disk center.    

\begin{figure}[h]
\includegraphics[scale=0.4]{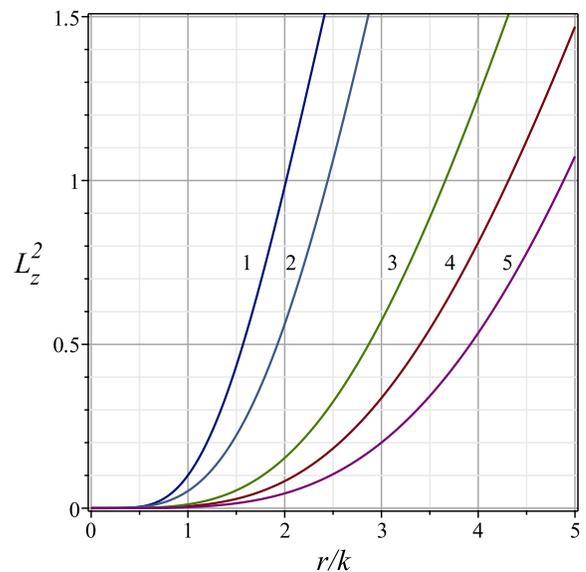}
\caption{The squared angular momentum in the $z$ direction $L^2_z$, given by (\ref{l2}), for a circular orbit on the disk, as a function of $r$. The curves correspond to the disk configurations enumerated
in Fig. \ref{conc}. It is clear that we have a monotonically increasing function of $r$ for all
curves. }
\label{Fig5}
\end{figure}

\begin{figure}[h]
\includegraphics[scale=0.4]{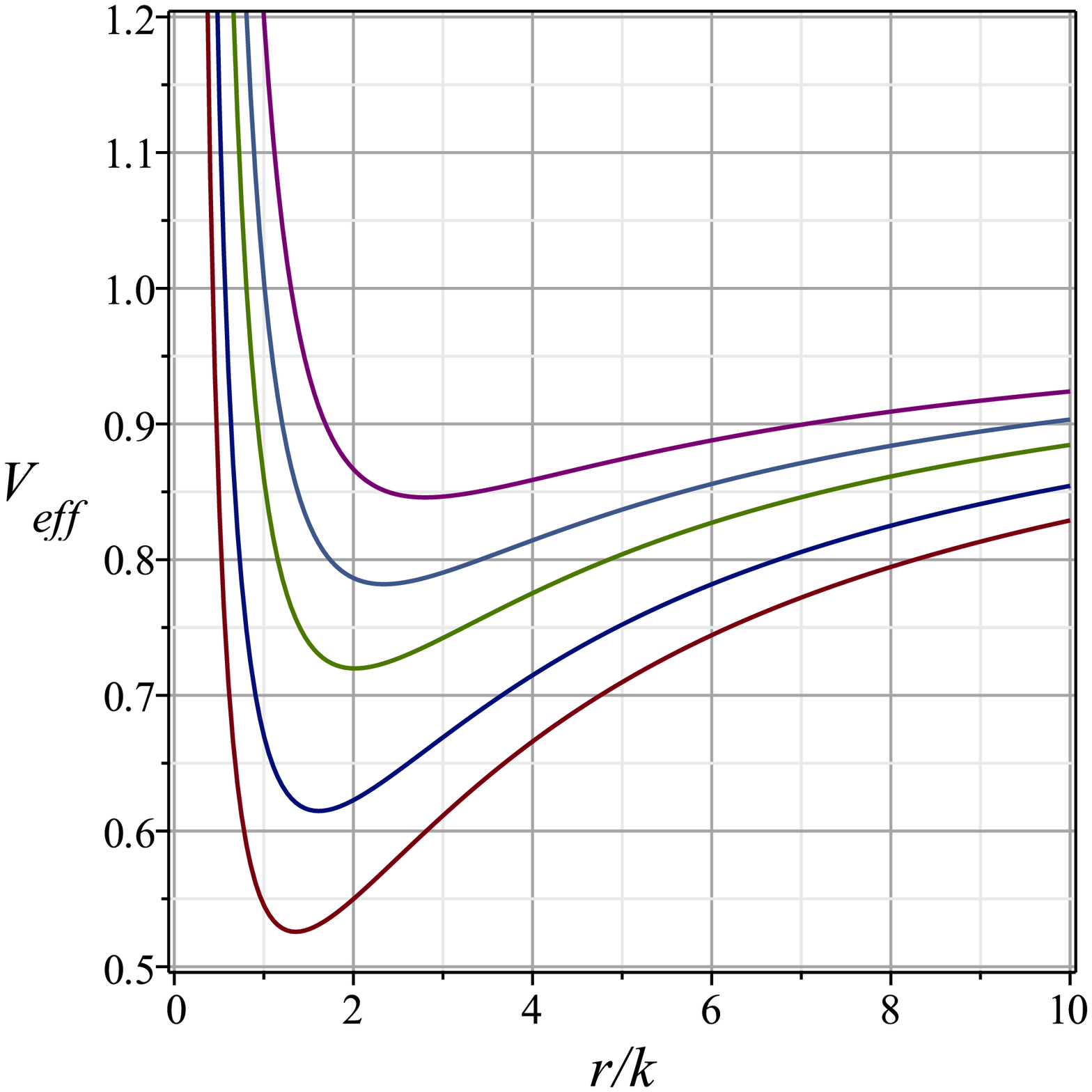}
\caption{The effective potential $V_{\rm eff}$ given by Eq. (\ref{Veff}) for the disk configurations of Fig. \ref{epsi}, with the curves disposed from top to bottom, and $L_z^2=1$. Other values for $L_z^2$ will only change the repulsive behavior near the origin. The circular orbits at $\bar r$ correspond to the minimum of  $V_{\rm eff}$. It is clear that we can have radially bounded motion around $\bar r$.  }
\label{Fig7}
\end{figure}

The Rayleigh criterion for radial stability  demands that both $e^\psi$ and  $L_z^2$ be monotonically increasing
functions of $r$ on the disk plane. Fig. \ref{Fig5} depicts the aspect of the
function  $L_z^2$  in the disk central region.
Since we have
$ L_z^2 \sim mr $ for large $r$, we conclude from Fig. \ref{Fig5} that the allowed configurations (see Fig. \ref{conc}) are indeed radially stable.  It is interesting also
to inspect the aspect of the effective radial potential (\ref{Veff}) on the disk center, see Fig. \ref{Fig7}. For large $r$, we have $V_{\rm eff}\sim 1-m/r$.  It is clear that we can have radially stable non-circular motion as well. 

Finally, one needs to check the vertical stability criterion, namely the positivity of  $\partial_{|z|}e^\psi$ on the disk plane. However, it turns out that the aspect of the 
function $\partial_{|z|}e^\psi$ is very similar to the surface density (\ref{Fig4}) depicted in
Fig. \ref{Fig4}, including the appearance of some maxima at $r_*>0$ for large values of
$\alpha$. Nevertheless, for the allowed configurations (those ones lying below the
red (solid) line in Fig. \ref{conc}), this function is always positive, assuring both the
radial and vertical stability of circular orbits in our disks.

\subsection{The disk surface currents}

Since our disks have   dipole magnetic fields, it is mandatory to investigate their origin. 
We will see that  the magnetic fields are supported entirely by surface polar  currents on the disk plane. Notice that for the electromagnetic quadripotential (\ref{quadri}), the nonvanishing components of the electromagnetic tensor are $F_{\varphi r} = - F_{r\varphi} = \partial_rA_\varphi$ and
 $F_{\varphi z} = - F_{z\varphi} = -\partial_zA_\varphi$. The former is associated with the magnetic field $B_z$ in the $z$ direction, while the latter is its radial component $B_r$. The application
 of (\ref{cdr}) without (\ref{cdr1}) would produce a discontinuous $B_z$ across the disk plane $z=0$, while $B_r$ would be continuous. Such kind of discontinuity would lead unavoidably  to a nonvanishing divergence for the magnetic field, implying the annoying presence of magnetic monopoles on the disk plane, see \cite{monopolos} for further details on this issue. By applying simultaneously (\ref{cdr}) and (\ref{cdr1}), the magnetic monopoles are avoided, since now the normal component of the magnetic field $B_z$ is continuous across the disk plane, while the discontinuity  is restricted to the radial part $B_r$. However, such a discontinuity in the parallel direction of the disk plane is not a problem at all, since it can be explained naturally due to the presence of superficial currents on the disk. In fact, the non-homogeneous Maxwell equation
 \begin{equation}
 \label{Max}
 \frac{1}{\sqrt{-g}}\partial_\mu\sqrt{-g} F^{\mu\nu} = 4\pi J^\nu
 \end{equation}
 can be invoked here to determine the surface current $j_\varphi$ such that
 \begin{equation}
 J^\nu = g^{\nu\varphi} j_\varphi \hat{\delta}(z),
 \end{equation}
 where $\hat\delta(z)$ stands for the covariant Dirac $\delta$-function. Applying the
 divergence theorem in (\ref{Max}) and taking into account that the parallel component
 of the magnetic field is discontinuous,  one has
 \begin{equation}
 j_\varphi = \frac{e^{-\eta/2}}{2\pi} \frac{\partial A_\varphi}{\partial |z|},
 \end{equation}
 leading to the invariant
 \begin{equation}
 \label{j2}
 j^2 = 
 j_\varphi j^\varphi = \frac{e^{\psi-\eta}}{4\pi^2 r^2} \left(\frac{\partial A_\varphi}{\partial |z|}\right)^2,
 \end{equation}
\begin{figure}[h]
\includegraphics[scale=0.4]{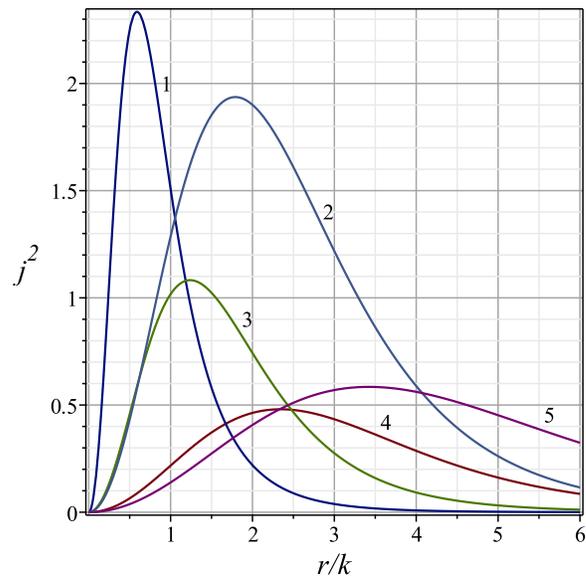}
\caption{The surface electric current invariant $j^2\,(\times 10^7)$  given by (\ref{j2}), as a function of $r$ for the allowed disk configurations of Fig. \ref{conc}. The curves corresponding to the configurations number 1 and 5 are re-scaled to fit in the graphics. They are, respectively,   multiplied by $10^{-2}$ and $10^2$.  }
\label{Fig6}
\end{figure}which aspect is illustrated in Fig. \ref{Fig6} for the disk configurations considered in Fig. \ref{conc}. 
Notice that, for large $r$, we have (see the Appendix for details on the asymptotic analysis) 
the following asymptotic behavior for the current invariant
\begin{equation}
\label{j2asympt}
j^2 \sim \frac{9\mu^2z_0^2}{4\pi^2r^8},
\end{equation}
implying that the total electromagnetic energy stored in the disk surface currents is finite. 
Also from Fig. \ref{Fig6}, one can see that the surface current is always distributed in 
a smooth ring-like structure, irrespective of the disk superficial density. The maximum of $j^2$ is typically located at a radius $r>0$ smaller than $z_0$, suggesting strongly that the   ring-like structure in the energy density profile of Fig. \ref{Fig4} has its origin
precisely in the surface currents.
 
\section{Final remarks}
\label{Conclusion}

Starting from the Gutsunaev-Manko solution \cite{GutsunaevManko1, GutsunaevManko2, GutsunaevManko3} describing massive magnetic dipoles in General Relativity, we have generated a continuous three-parameter family of solutions corresponding
to static magnetized thin disks. We have adapted 
  the well known  ``displace, cut and reflect'' DCR  method, due to Kuzmin \cite{DCR},
to avoid the presence of magnetic monopoles on the disk. Essentially, we combine with the
usual steps of the DCR method the reflection (\ref{cdr1}) on the electromagnetic potential. In this way, we obtain a field configuration such that the magnetic field component $B_z$ perpendicular  to the disk is continuous, whereas the parallel component $B_r$ is discontinuous, leading to a physical situation without magnetic monopoles, but with the magnetic field entirely supported by surface polar currents. Moreover,  all disk solutions such that the parameters $(z_0,\alpha)$ are in the allowed region (laying below the red (solid) curve in Fig. \ref{conc}) have circular obits stable against radial and vertical perturbations.
Since our disks have no radial pressure, it can be considered as formed by counter-rotating identical particles, and them the stability of circular orbits is essential to establish the stability of the solution. 

A certainly pertinent question here is if the two stability test we have performed would be enough to assure the stability of the disk beyond the counter-rotating hypothesis.  This is quite complicated problem. Since the matter content of the disk has azimuthal pressure, the motion of its constituents will not be purely geodesic as one would expect, for instance, for real dust disks (no pressure at all). The stability analysis of the 
disk fluid does require extra information like, for instance, the fluid equation of state, see  \cite{Max}. Also, the dynamics of oblique orbits in disks with central fields is known to be generically  chaotic \cite{SaaVenegeroles,Saa,Hunter1,Hunter2,Chaos1,Chaos2,Chaos3,Chaos4}, challenging the view that the counter-rotating particles will keep their circular orbits for long times. These are more subtle question that we can now put forward once we have established that our magnetized disks pass by the simplest stability tests. Our results are, in this sense, the first step to a   deeper study of the stability of these disks. 

Finally, we have restricted the application of the DCR method   for the cases where $z_0>k$. In order to enlighten such an option, let us consider the inverse of the transformations (\ref{t1}) and (\ref{t2}), namely
\begin{eqnarray}
r &=& k \sqrt{x^2-1}\sqrt{1-y^2},\\
z &=& kxy. 
\end{eqnarray}
Recalling that the prolate coordinates are such that  $x\ge 1$  and $-1 \le y \le 1$,  we see that the vertical segment $|z|\le k$ at the origin $r=0$ corresponds to $x=1$ and $|y|<1$. However, for $x=1$ we have $e^\psi=0$, see (\ref{gs1}), and so any disk constructed by choosing the hyperplane $z_0 < k$ will unavoidably encounter the complicated, and not yet quite well understood, causal structure of the central part of the Gutsunaev-Manko solution. In particular, it would be impossible to attain any stable configuration. The situation is
analogous to the disks generated from Schwarzschild solution. Nevertheless, disk distributions approaching the horizon of generic black holes are certainly interesting for the study of accretion disks\cite{Semerak,Karas,Polish}. This is a rather promising possibility for our magnetized disks that would deserve further investigations since they might give some insights of possible observational signatures which one could seek, for instance, with the Event Horizon Telescope\cite{EHT}.

\subsection*{Acknowledgments}

The authors are grateful to CNPq, FAPERJ (grant E-26/200.279/2015, VPF)  and FAPESP (grant 2013/09357-9, AS)  for the financial support,
and R.S.S. Vieira and J. Ramos-Caro for enlightening discussions on thin disks in General
Relativity, and an anonymous referee for comments and suggestions that improved the paper. 

\appendix

\section*{Appendix}

The mathematical expressions in this work involve  essentially only algebraic rational  functions and, hence, 
the evaluation of derivatives and the location of zeros could, in principle, be straightforwardly done. Nevertheless, the resulting expressions are typically huge, and we have indeed 
taken advantage of Maple Software to deal with them.  However, we have some useful simplifications for the asymptotic analysis of large $r$ and in the central region of the disk. 
The calculation of the disk central density (\ref{sigma0}), for instance, can  be considerably
simplified noticing that, at the disk center, which has prolate coordinates $x=\frac{z_0}{k}$ and $y=1$,
one has  $g=1$, $\frac{\partial x}{\partial |z|}  = \frac{1}{k}$, and  $\frac{\partial y}{\partial |z|}  = 0$.
 The other relevant quantities at the disk center are
\begin{eqnarray}
e^\psi &=& e^{-\eta} = \frac{z_0-k}{z_0+k}f^2, \\
\frac{\partial e^\psi}{\partial |z|} &=& \frac{2kf^2}{(z_0+k)^2} + \frac{2f}{k}\frac{z_0-k}{z_0+k}\frac{\partial f}{\partial x}  ,
\end{eqnarray}
and
\begin{equation}
\frac{\partial e^\eta}{\partial |z|} = \frac{1}{f^{2}}\left( \frac{1}{k} \frac{z_0+k}{z_0-k}\left(\frac{\partial g^2}{\partial x} - \frac{2}{f}\frac{\partial f^2}{\partial x}\right) -\frac{2k}{(z_0-k)^2}\right).
\end{equation}
It turns out that $\partial_xg^2$ vanishes at the disk center, as one can check after some algebra. 
Combining these results leads to 
\begin{equation}
e^\eta\frac{\partial e^\psi}{\partial |z|} -e^\psi\frac{\partial e^\eta}{\partial |z|}  =
\frac{4k}{z_0^2-k^2} + \frac{4}{kf}\frac{\partial f}{\partial x},
\end{equation}
from where  (\ref{sigma0}) follows directly. The values of $f$ and $f_x$ at the disk center, necessary to determine the zeros of (\ref{sigma0}), are
\begin{equation}
f^{(0)} = \frac{\left(1+\alpha^2\right)^2\left(z_0^2 - k^2 \right)^2}{\left(z_0^2-k^2+\alpha^2(z_0-k)^2\right)^2-4\alpha^2k^2(z_0^2-k^2)}, 
\end{equation}
and
\begin{equation}
f_x^{(0)} =4\left( z_0 - \frac{1}{k^2}h^{(0)} f^{(0)}\right)f^{(0)},
\end{equation}
where
\begin{equation}
h^{(0)} =  (z_0-k)^3\alpha^4 + (2z_0^3 -3kz_0^2 -2k^2z_0 +k^3)\alpha^2 + z_0^3 -k^2z_0  .
\end{equation}

The concavity analysis of $e^\psi$ in Fig. \ref{conc} requires the evaluation of 
$\partial_r e^\psi$ and $\partial_r^2 e^\psi$ at the disk center, $r=z=0$, or 
$x=\frac{z_0}{k}$ and $y=1$. Since $\frac{\partial x}{\partial r}  =\frac{\partial y}{\partial r}  = 0$ at the disk center, we will also have $\partial_r e^\psi =0$. For the second
derivative, we have 
\begin{equation}
\frac{\partial^2 x}{\partial r^2} = \frac{z_0/k}{z_0^2-k^2}
\end{equation}
and
\begin{equation}
\frac{\partial^2 y}{\partial r^2} = -\frac{1}{z_0^2-k^2}
\end{equation}
at $r=z=0$, leading to
\begin{equation}
\partial_r^2 e^\psi = \frac{2f}{z_0^2-k^2} \left( \frac{z_0^2}{k^2} \frac{f}{(z_0+k)^2} 
+  \frac{z_0-k}{z_0+k}\left(
\frac{z_0}{k}  \partial_x f
-   \partial_y f
\right)\right),
\end{equation}
which can be calculated analogously to the case of $\frac{\partial e^\psi}{\partial |z|}$
above. The critical values $\alpha^2_*$ of Fig. \ref{conc} are the roots of the higher order polynomial corresponding to $\partial_r^2 e^\psi=0$ at $r=z=0$.

For the asymptotic behavior of our solutions for large $r$ on the disk plane $z=0$, notice that from (\ref{t1}) and (\ref{t2}), we have that $x\sim \frac{r}{k}$ and $y\sim \frac{z_0}{r}$  for $z=0$ and large $r$. For their derivatives, we
have
\begin{eqnarray}
\frac{\partial x}{\partial r}&\sim & \frac{1}{k}, \quad\frac{\partial x}{\partial z}\sim \frac{z_0}{kr}, \\
\frac{\partial y}{\partial r}&\sim & \frac{z_0}{r^2}, \quad\frac{\partial y}{\partial z}\sim \frac{1}{r}, 
\end{eqnarray}
in the same asymptotic limit. From these asymptotic limits, one can get  
\begin{eqnarray}
\label{asy1}
&\displaystyle e^\psi  \sim 1 - \frac{2m}{r},\quad  e^\eta  \sim 1 + \frac{2m}{r} &\\
\label{asy2}
&\displaystyle \frac{\partial e^\psi}{\partial r} \sim -\frac{\partial e^\eta}{\partial r} \sim \frac{2m}{r^2}& \\
\label{asy3}
&\displaystyle \frac{\partial e^\psi}{\partial |z|} \sim - \frac{\partial e^\eta}{\partial |z|} \sim  \frac{2mz_0}{r^3}&
\end{eqnarray}
with $m$ given by (\ref{mass}). For the magnetic potential (\ref{quadri2}), we have
\begin{equation}
\frac{\partial A_\varphi}{\partial |z|} \sim -\frac{24\alpha^3 }{(1+\alpha^2)^2}
\frac{z_0k^2}{r^3} = -
\frac{3\mu z_0}{r^3},
\end{equation} 
for large $r$ at $z=0$, 
with $\mu$ given by (\ref{mu}), from where (\ref{j2asympt}) follows directly.

\end{document}